\documentclass[11pt,twoside]{article}
\usepackage{acta-i,euler}
\newcommand{\uj}{\bigskip\noindent}

\newcommand{\vol}{\textbf{1}, 1 (2009) 109--118}   
\newcommand{\amss}{\amssubj{68R05
}}     
{{
}}   
\newcommand{\CRs}{\CRsubj{%
G.2.1. [Combinatorics]: Subtopic - Combinatorial algorithms.
}}   
\newcommand{\keyw}{\keywords{%
 Dyck words, generating and ranking algorithms, Catalan numbers
}}

\newcommand{\key}[1]{\textbf{#1}}

\newcommand{\tsc}[1]{\textrm{\textsc{#1}}}

\newcommand{\alge}{%
\begin{tabbing}%
99 \= xxx\=xxx\=xxx\=xxx\=xxx\=xxx\=xxx\=xxx\=xxx\=xxx\=xxx \+ \kill 
}

\newcommand{\algeN}{%
\begin{tabbing}%
\hspace*{1pt}999\=xxx\=xxx\=xxx\=xxx\=xxx\=xxx\=xxx\=xxx\=xxx\=xxx\=xxx \+ \kill  
}

\newcommand{\algv}{%
\end{tabbing}
}

\newenvironment{algN}[1]{%
\vspace{4mm}  
\vbox\bgroup\noindent\tsc{#1}%
\vspace*{-2mm}  
\algeN}
{
\algv\egroup
\vspace{0mm}  
}

\title{Generating and ranking of Dyck words}

\short{Z. K\'asa}{Generating and ranking of Dyck words}

\begin{document}

\maketitle

\oneauthor{Zolt\'an K\'asa
}{Sapientia -- Hungarian University of Transylvania\\
Department of Mathematics and Informatics, T\^argu Mure\c s
}{kasa@ms.sapientia.ro
}

\begin{abstract}
A new algorithms to generate all Dyck words is presented, which is used in ranking and unranking Dyck words. We emphasize the importance of using Dyck words in encoding objects related to Catalan numbers.  As a consequence of formulas used in the  ranking algorithm we can obtain a recursive formula for the $n$th Catalan number.
\end{abstract}

\section{Introduction}

Let $B=\{ 0,1\}$ be a binary alphabet and $x_1x_2\ldots x_{n}\in B^{n}$. Let $h:B\rightarrow \{-1, 1\}$ be a valuation function with $h(0)=1$, $h(1)=-1$, and  $h(x_1x_2\ldots x_{n})= \displaystyle\sum_{i=1}^{n}{h(x_i)}$.

A word $x_1x_2\ldots x_{2n}\in B^{2n}$ is called a \emph{Dyck word} \cite{duchon} if it satisfy the following conditions:
\begin{eqnarray*}
 &&  h(x_1x_2\ldots x_{i})\ge 0, \textrm{ for } 1\le i\le 2n-1\\
 &&  h(x_1x_2\ldots x_{2n})=0.
\end{eqnarray*}
$n$ is the semilength of the word.

\section{Lexicographic order}

The algorithm that generates all Dyck words in lexicographic order is obvious. Let us begin with 0 in the first position, and add 0 or 1 each time the Dyck-property remains valid. In the following algorithm $2n$ is the length of a Dyck word, $n_0$ counts the 0s, and $n_1$ the 1s.

\noindent There are the following cases :

{\small  \noindent Case 1:  ($n_0<n$) and ($n_1<n$) and  ($n_0>n_1$)  \hfill (We can continue by adding 0 and 1.)
   
\noindent   Case 2:  ($n0<n$) and ($n_1<n$) and  ($n_0=n_1$)   \hfill (We can continue by adding 0 only.)
   
\noindent   Case 3:  ($n_0<n$) and ($n_1=n$) \hfill (We can continue by adding 0 only.)
   
 \noindent  Case 4:  ($n_0=n$) and ($n_1<n$) \hfill (We can continue by adding  1 only.)
   
 \noindent  Case 5:  ($n_0=n_1=n$) \hfill (A Dyck word is obtained.) 
}

\noindent Let us use the following short notations:

\medskip
\begin{minipage}[t]{5cm}
 \begin{tabbing}
\emph{Dyck 0} for \\
\qquad \= $ x_i:=0$\\
        \>  $n_0:=n_0+1$\\
        \>  \textsc{LexDyckWords}($X,i,n_0,n_1)$\\
        \> $n_0:=n_0-1$
\end{tabbing}
\end{minipage}\qquad
\begin{minipage}[t]{5cm}
\begin{tabbing}
\emph{Dyck 1} for \\
\qquad  \= $ x_i:=1$\\
        \>  $n_1:=n_1+1$\\
        \>  \textsc{LexDyckWords}($X,i,n_0,n_1)$\\
        \> $n_1:=n_1-1$
\end{tabbing}
\end{minipage}

\uj The algorithm is the following:

\begin{algN}{LexDyckWords($X,i,n_0,n_1$)}
1 \' \key{if}  \= Case 1  \\
2 \'           \>  \key{then} \=  $i:=i+1$\\
3 \'           \>         \> \emph{Dyck 0}\\
4 \'           \>         \> \emph{Dyck 1}\\
5 \' \key{if}   Case 2 \key{or} Case3 \\
6 \'           \>  \key{then} $i:=i+1$\\
7 \'           \>    \> \emph{Dyck 0}\\ 
8 \' \key{if}  \= Case 4  \\
9 \'           \>  \key{then} $i:=i+1$\\ 
10 \'           \> \>  \emph{Dyck 1} \\
11 \' \key{if}  Case 5  \\
12 \'           \>  \key{then}  Visit  $x_1x_2 \ldots x_n$ \\
13 \'  \key{return}
\end{algN}

\noindent The recursive call:

  $x_1=0$, $n_0:=1$, $n_1:=0$
  
  \textsc{LexDyckWords}($X,1,n_0,n_1$)

\uj
For $n=4$ the following result:

\medskip
00001111, 00010111, 00011011, 00011101, 00100111, 00101011, 00101101,

00110011, 00110101, 01000111, 01001011, 01001101, 01010011, 01010101.

\uj This algorithm obviously generates all Dyck words.

\section{Generating the positions of 1s}

Let $b_1b_2\ldots b_{n}$  be the positions of 1s in the Dyck word $x_1x_2\ldots x_{2n}$. E.g. for $x_1x_2\ldots x_8=01010011$ we have
 $b_1b_2b_3b_4=2478.$ 

To be a Dyck word of semilength $n$, the positions $b_1b_2\ldots b_{n}$ of 1s of the word $x_1x_2\ldots x_{2n}$ must satisfy the following conditions:
\[   2i\le b_i \le n+i, \quad\textrm{ for } \quad 1\le i \le n.\]

Following the idea of generating combinations by positions of 0s in the corresponding binary string \cite{knuth3} we 
propose a similar algorithm that generates the positions $b_1b_2\ldots b_{n}$ of 1s.

\begin{algN}{PosDyckWords($n$)}
1 \' \key{for} \= $i:=1$ \key{to} $n$ \\
2 \'           \>  \key{do} $b_i := 2i$ \\
3 \'  \key{repeat} \\
4 \'         \>  Visit  $b_1 b_2 \ldots b_n$\\
5 \'         \> $IND:=0$\\
6 \'         \> \key{for} \= $i:=n-1$ \key{downto} 1 \\
7 \'         \>           \> \key{do} \= \key{if} \= $b_i < n+i$ \\
8 \'         \>           \>          \>          \> \key{then} \= $b_i:=b_i+1$\\
9 \'         \>           \>          \>          \>            \> \key{for} \= $j:= i+1$ \key{to} $n-1$\\
10 \'         \>           \>          \>          \>            \>   \> \key{do} \= $b_j:=\max (b_{j-1}+1, 2j)$\\
11 \'         \>           \>          \>          \>            \>  $IND:=1$\\
12 \'         \>           \>          \>          \>            \>  \key{break (for)}\\
13 \'  \key{until} $IND=0$\\
14 \'  \key{return}
\end{algN}

\uj  For $n=4$ the following  result:

\uj 2468, 2478, 2568, 2578, 2678, 3468, 3478, 
3568, 3578, 3678, 4568, 4578, 4678, 5678

\uj The corresponding Dyck words are:

\medskip 01010101, 01010011, 01001101, 01001011, 01000111,  00110101, 00110011,

00101101, 00101011, 00100111, 00011101, 00011011, 00010111, 00001111

\uj Because all values of positions that are possible are taken by the algorithm, it generates all Dyck words. Words are generated in reverse lexicographic order.

\section{Generating by changing 10 in 01}
The basic idea \cite{bege2} is to change the first occurence of 10 in  01 to get a new Dyck word. We begin with $0101\ldots 01$.
 
\uj Let us denote by $X$ the Dyck word $x_1x_2\ldots x_{2n}$. 
 
\uj
\begin{algN}{DyckWords($X,k$)}
1 \' $i := k$\\
2 \' \key{while} \= $i < 2n$\\ 
3 \'          \> \key{do} \= Let $j$ \= be the position of the first occurence of 10 in $x_ix_{i+1} \ldots x_{2n}$,\\ 
               \>          \>            or 0 if such a position doesn't exist.\\
4 \'               \>              \> \key{if} \= $j>0$\\
5 \'               \>              \>     \> \key{then} \= Let $Y:=X$\\
6 \'               \>              \>     \>            \> Change $y_i$ with $y_{i+1}.$\\ 
7 \'                \>              \>    \>              \> Visit $y_1y_2 \ldots y_{2n}$\\
8 \'                \>              \>    \>              \> \textsc{DyckWords}($Y,j-1$)\\
9 \'                \>              \>    \>              \> $i:=j+2$\\
10 \'  \key{return}
\end{algN}
The first call is  {\textsc{DyckWords}($X, 1$)},  if $X=0101\ldots 01$.

\uj For  $X=01010101$, the algoritm generates:

\medskip
01010101, 00110101, 00101101, 00011101, 00011011, 00010111, 00001111,

00101011, 00100111, 00110011, 01001101, 01001011, 01000111, 01010011.

\uj Can this algorithm always generate all Dyck words? To prove this we show that any Dyck word can be tranformed to $(01)^n$ by several changing of 01 in 10. Let us consider the leftmost subword of the form $0^i1$, for $i>0$. Changing 01 in 10 $(i-1)$ times, we will obtain a leftmost subword of the form $0^{i-1}1$. So, all subwords of this form can be avoided.

\section{Ranking Dyck words}

Ranking Dyck words means \cite{liebe} to determine the position of a Dyck word in a given ordered sequence of all Dyck words.

Algorithm \textsc{PosDyckWords} generates all Dyck word in reverse lexicographic order. For ranking these words we will use the following function \cite{wuu}, where $f(i,j)$ represents the number of paths  between (0,0) and $(i,j)$ not crossing the diagonal $x=y$ of the grid.

\begin{equation}
f(i,j)= \left\{ \begin{array}{ll}
                1,                 & \textrm{for } 0\le i \le n, j=0\\
                f(i-1,j)+f(i,j-1), & \textrm{for } 1\le j < i\le n\\
                f(i,i-1),          & \textrm{for } 1\le i=j \le n\\
                0,                 & \textrm{for }  0\le i <j\le n  
                \end{array} \right.
\label{egy}
\end{equation}
Some values of this function are given in the following table.

\uj
\begin{center}
\begin{tabular}{r|rrrrrrrrrrr}
9 &   &   &   &   &   &   &   & &&4862&\\
8 &   &   &   &   &   &   &   & &1430&4862&\\
7 &   &   &   &   &   &   &   &429&1430&3432&\\
6 &   &   &   &   &   &   &132&429&1001&2002&\\
5 &   &   &   &   &   & 42&132&297&572&1001&\\
4 &   &   &   &   & 14& 42& 90&165&275&429&\\
3 &   &   &   & 5 & 14& 28& 48& 75&110&154&\\
2 &   &   & 2 & 5 & 9 & 14& 20& 27&35&44&\\
1 &   & 1 & 2 & 3 & 4 & 5 & 6 & 7 &8&9&\\
0 & 1 & 1 & 1 & 1 & 1 & 1 & 1 & 1 &1&1&\\ \hline
$\uparrow$ & 0 & 1 & 2 & 3 & 4 & 5 & 6 & 7 & 8&9&$\leftarrow i$\\
$j$  &   &   &   &   &   &   &   & &\\ 
\end{tabular}
\end{center}

\uj
It is easy to prove that if $C_n$ is the $n$th Catalan number then

\begin{equation}
C_{n+1}= f(n+1,n)=\sum_{i=0}^{n}{f(n,i)}, \quad n\ge 0
\label{ketto}
\end{equation}
\[ f(n+1,k)=\sum_{i=0}^{k}{f(n,i)}, \quad n\ge 0, n\ge k\ge 0. \]
Using this function the following ranking algorithm results.

\uj\begin{algN}{Ranking($b_1b_2\ldots b_n$)}
1 \' $c_1:=2$\\
2 \' \key{for} \= $j:= 2$ \key{to} $n$ \\ 
3 \'            \> \key{do} \= $c_j:= \max(b_{j-1}+1, 2j) $\\
4 \' $nr:= 1$    \\                    
5 \' \key{for} $i:=1$ \key{to} $n-1$\\  
6 \'            \> \key{do} \key{for} \= $j:= c_i$ \key{to} $b_i-1$  \\
7 \'            \>                    \>  \key{do} $nr:=nr+f(n-i,n+i-j)$\\
8 \'  \key{return} $nr$
\end{algN}
For example, if $b$ =4 5 8 9 10, we get $c$ =2 5 6 9 10, and 
$nr= 1+f(4,4)+f(4,3)+f(2,2)+f(2,1)= 1+14+14+2+2=33.$

\noindent This algorithm can be used for ranking in lexicographic order too.

\section{Unranking Dyck words}

The unranking algorithm for a given $n$ will map a number between 1 and $C_n$ to the corresponding Dyck word represented by positions of 1s. Here the Dyck words are considered in reverse lexicographic order too.

\uj\begin{algN}{Unranking($nr$)}
1 \' $b_0:=0$\\
2 \' $nr:=nr-1$\\
3 \' \key{for} \= $i:= 1$ \key{to} $n$ \\ 
4 \'           \> \key{do} \=  $b_i:= \max \big(b_{i-1}+1, 2i\big)$  \\                    
5 \'            \>         \>  $j:=n+i-b_i$ \\  
6 \'            \>         \>  \key{while} \= \big($nr \ge f(n-i,j)$\big) \key{and} \big($b_i < n+i$\big)  \\
7 \'            \>         \>              \>  \key{do} \= $nr:=nr-f(n-i,j)$\\
8 \'            \>         \>              \>           \> $b_i:=b_i+1$ \\
9 \'            \>         \>              \>           \> $j:=j-1$\\
10 \'  \key{return} $b_1b_2\ldots b_n$
\end{algN}
If $n=6$ and $nr=93$, we will have: $92-f(5,5)-f(5,4)-f(3,3)-f(2,2)-f(1,1)= 92-42-42-5-2-1$, so the corresponding Dyck word represented by positions of 1's is: $b$ = 4 5 7 9 11 12. Are changed from the initial values $2i$ the following:  position 1 by 2, position 3 by 1, position 4 by 1 and position 5 by 1.

\section{Applications of Dyck words}

If ${\cal O}$ is a set of $C_n$ objects, Dyck words can be used for encoding the objects of ${\cal O}$. 
The importance of such an encoding currently is not suitably accentuated. We present here an encoding and decoding algorithms for binary trees, based on \cite{bege}.  

\uj\emph{Algorithm for encoding a binary tree}

\noindent Let $B_L$ be the left and $B_R$ the right subtree of the binary tree $B$.  $w01$ means the concatenation of word $w$ with 01, and $w$ is considered a global variable.

\begin{algN}{EncodingBT($B$)}
1 \' \key{if} \= $B_L \ne \emptyset$ and $B_R = \emptyset$ \\
2 \'           \>  \key{then} \= $w:=w01$ \\
3 \'            \>            \> \textsc{EncodingBT}($B_L$)  \\
4 \' \key{if} \= $B_L = \emptyset$ and $B_R \ne \emptyset$ \\
5 \'           \>  \key{then} \= $w:=w10$ \\
6 \'            \>            \> \textsc{EncodingBT}($B_R$)  \\
7 \' \key{if} \= $B_L \ne \emptyset$ and $B_R \ne \emptyset$ \\
8 \'           \>  \key{then} \= $w:=w00$ \\
9 \'            \>            \> \textsc{EncodingBT}($B_L$)  \\
8 \'           \>             \> $w:=w11$ \\
9 \'            \>            \> \textsc{EncodingBT} ($B_R$)  \\
10 \' \key{return} 
\end{algN}

\uj Call:  

 $w:= 0$

 \textsc{EncodingBT}($B$)

 $w:=w1$

\uj  For all trees of $n=4$ vertices the result of the algorithm is given in Fig. 1. 

\begin{figure}[t]
\begin{center}
\includegraphics[scale=0.6]{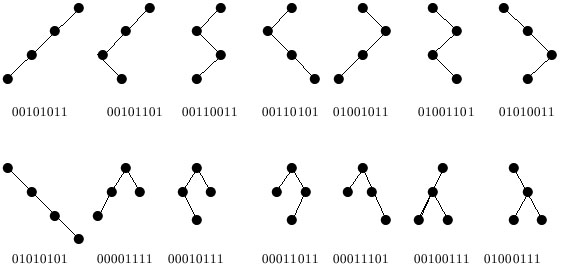}
\end{center}
\caption{Encoding of binary trees for $n=4$.}
\label{Fig1}
\end{figure}

\uj\emph{Algorithm to decode a Dyck word into a binary tree}

\noindent At the beginning the root of the generated binary tree is the current vertex. When an edge is drawn, its endvertex becomes the current vertex. 

\newpage
\begin{algN}{DecodingBT($w$)}
1 \' Let  $ab$ be the first two letters of $w$.\\
2 \' Delete $ab$ from $w$.\\
3 \' \key{if} \= $ab = 01$ \\
4 \'          \> \key{then} \= draw a left edge from the current vertex\\
5 \'          \>            \> \textsc{DecodingBT}($w$)\\
6 \' \key{if} \= $ab = 10$ \\
7 \'          \> \key{then} \= draw a right edge from the current vertex\\
8 \'          \>            \> \textsc{DecodingBT}($w$)\\
9 \' \key{if} \= $ab = 00$ \\
10 \'          \> \key{then} \= put in the stack the position of the current vertex\\
11 \'          \>            \> draw a left edge from the current vertex\\
12 \'          \>            \> \textsc{DecodingBT}($w$)\\
13 \' \key{if} \= $ab = 11$ \\
14 \'          \> \key{then} \= get from the stack the position of the new current vertex\\
15 \'          \>            \> draw a right edge from the current vertex\\
16 \'          \>            \> \textsc{DecodingBT}($w$)\\
17 \' \key{return} 
\end{algN}

\noindent Call:

  delete 0 from the beginning and 1 from the end of the input word $w$ 

  draw a vertex (the root of the tree) as current vertex

  \textsc{DecodingBT}($w$)

\uj For some other objects related to Catalan numbers the corresponding coding can be found in \cite{bege} and at
http://www.ms.sapientia.ro/\~{}kasa/CodingDyck.pdf. 

\section{A consequence}

As a consequence of formulas (\ref{egy}) and (\ref{ketto}) the following formula for the $(n+1)$th Catalan number results:

\begin{equation}
C_{n+1}= 1+ \sum_{k\ge0}^{} {(-1)^k\binom{n-k}{k+1}C_{n-k}}. 
\label{harom}
\end{equation}

\noindent We can prove that 
\[ f(n,n-k)= \sum_{i=0}^{n}{(-1)^i\binom{k-i}{i}C_{n-i}}\] 
for appropriate $n$ and $k$, using mathematical induction on $n$ and $k$,   and formula (\ref{egy}) in the form 
\[f(n,n-k)= f(n,n-k+1)-f(n-1, n-k+1).\]

\noindent  Now, from (\ref{ketto}) 
\begin{eqnarray*}
C_{n+1} &=& \sum_{i=0}^{n}{f(n,i)}=f(n,0) + \sum_{i=1}^{n}{f(n,i)}=1+ \sum_{i=0}^{n-1}{f(n,n-i)}\\
        &=&  1+ \sum_{i=0}^{n-1}{\left( \sum_{k=0}^{n}{(-1)^k \binom{i-k}{k}C_{n-k}}\right)}\\
        &=&  1+ \sum_{k=0}^{n}{(-1)^kC_{n-k}\left( \sum_{i=0}^{n-1}{\binom{i-k}{k}}\right)}\\ 
        &=&  1+ \sum_{k=0}^{n}{(-1)^k\binom{n-k}{k+1}C_{n-k}}.
\end{eqnarray*}
In the last line $\binom{k}{k}+  \binom{k+1}{k}+ \cdots +   \binom{n-1-k}{k}=  \binom{n-k}{k+1}$ has been used.

\bigskip
\rightline{\emph{Received: November 17, 2008}}     


\newpage
\begin{thebibliography}{9}

\bibitem{bege} A. Bege, Z. K\'asa,  Coding objects related to Catalan numbers, \emph{Studia Universitatis Babe\c s-Bolyai, Informatica} \textbf{46}, 1 (2001) 31--40.

\bibitem{bege2} A. Bege, Z. K\'asa, \textit{Algoritmikus kombinatorika \'es sz\'amelm\'elet}, Presa Universitar\u a Clujean\u a, 2006. 

\bibitem{deutsch} E. Deutsch,  Dyck path enumeration, \emph{Discrete Mathematics} \textbf{204}, 1--3 (1999) 167--202.


\bibitem{duchon} P. Duchon,  On the enumeration and generation of generalized Dyck words, \emph{Discrete Mathematics} \textbf{225}, 1--3 (2000) 121--135.


\bibitem{knuth3} D. E. Knuth, \textit{The Art of Computer Programming, Vol. 4. Fasc. 3, Generating All Combinations and Partitions}, Addison-Wesley, 2005. 


\bibitem{liebe} J. Liebehenschel, Ranking and unranking of lexicographically ordered words: An average-case analysis. \textit{Journal of Automata, Languages and Combinatorics}, \textbf{2}, 4 (1997) 227--268

\bibitem{wuu} W. Yang, \textit{Discrete Mathematics}, { http://www.cis.nctu.edu.tw/\~{}wuuyang/},  manuscript.

\end{thebibliography}
\end{document}